# Doppler Effect Associated with the Reflection of Light on a Moving Mirror

**Bernhard Rothenstein and Ioan Damian**

Politehnnica University of Timisoara, Department of Physics,
Timisoara, Romania
**bernhard_rothenstein@yahoo.com, ijdamian@yahoo.com**

**Abstract**
The Doppler Effect associated with the reflection on a moving mirror is reduced to two Doppler Effect experiments involving the incoming incident ray and the outgoing reflected ray or vice-versa. The dependence of the corresponding Doppler factors on the incidence angle on the stationary mirror.

The reflection of a plane-polarized light wave reflected from a uniformly moving mirror is associated with two Doppler Effect experiments. Figure 1 shows a vertical plane mirror, an incident and a reflected ray as detected from the mirror's rest frame K'(X'O'Y'). The incidence and the reflection take place at the origin O'(0.0) at a zero time $t'=0$ generating the event $E'_0(x'=0, y'=0, t'=0)$ that is characterized by the same space-time coordinates in all inertial reference frames in relative motion. The mirror and its rest frame move with constant velocity $V$ relative to the K(XOY) reference frame in the positive direction of the common OX(O'X') axes. The corresponding axes of the two frames are parallel to each other. At the origin of time in the two frames ($t=t'=0$) the origins O and O' are located at the same point in space.

The first Doppler Effect experiment involves the incident ray, observers $O'_0(0,0)$ and $O'_1(R'_1, \alpha)$ of the K' frame, $O_0(0,0)$ and $O_1(R_1, i)$ of the K frame. Let $C'_0(0,0)$; $C'_1(R'_1, \alpha)$; $C_0(0,0)$; $C_1(R_1, i)$ be theirs wrist watches, $i$ representing the incidence angle as measured from K. It differs from $\alpha$ due to the aberration of light effect.[1] We sketch it in Figure 2 and we derive the corresponding Doppler shift formula following a strategy proposed by Peres[2] which reveals the limits of its applicability and the fact that we can derive it without using the Lorentz-Einstein transformations.

When clock $C_1(R_1, i)$ reads $t_1$ observer $O(R_1, i)$ detects a wave crest of the incident ray. The same wave crest is detected by observer $O_0(0,0)$ when his clock $C_0(0,0)$ reads $t_0$ and it is obvious that

$$t_0 = t_1 + \frac{R_1}{c}. \qquad (1)$$

We allow small changes in the readings of the clocks, related by

$$dt_0 = dt_1 + \frac{dR_1}{c} \qquad (2)$$

an equation that holds for each value of $t_0$ even for a zero value. The clock $C_0'(0,0)$ of observer $O_0'(0,0)$ registers a change of its reading $dt_0'$ related to $dt_0$ by the time dilation formula

$$dt_0 = \frac{dt_0'}{\sqrt{1 - \frac{V^2}{c^2}}}. \qquad (3)$$

Presenting (2) as

$$1 = \frac{dt_1}{dt_0} + c^{-1} \frac{dt_1}{dt_0} \qquad (4)$$

and taking into account that by definition

$$\frac{dR_1}{dt_0} = V \cos i \qquad (5)$$

equation (4) leads to

$$\frac{dt_1}{dt_0'} = \frac{1 - \frac{V}{c} \cos i}{\sqrt{1 - \frac{V^2}{c^2}}}. \qquad (6)$$

We can consider that $dt_1$ and $dt_0'$ represent the periods of the electromagnetic oscillations taking place in the incident ray as measured in K and K' respectively.

The second Doppler Effect experiment is sketched in Figure 3 and we consider it from K'. When clock $C_0'(0.0)$ reads $t_0'$ observer $O_0'$ detects a wave crest of the reflected ray.

Observer $O_2'$ detects the same wave crest when his clock $C_2'$ reads $t_2'$ and it is obvious that

$$t_2' = t_0' + \frac{R_2'}{c}. \qquad (7)$$

Following the same strategy as in the first case we obtain

$$\frac{dt_2}{dt_0'} = \frac{1 + \frac{V}{c} \cos r}{\sqrt{1 - \frac{V^2}{c^2}}}. \qquad (8)$$

Eliminating $dt_0'$

between (6) and (8) we obtain

$$\frac{dt_2}{dt_1} = \frac{1 + \frac{V}{c} \cos r}{1 - \frac{V}{c} \cos i} \qquad (9)$$

recovering a result obtained by Rosser[3]. It offers little insight in the way in which $dt_2$ depends on $dt_1$ because (9) presents the relationship between them as a function of $i$ and $r$. There are two possible ways to obtain more insight. In the first case we can express $r$ as a function of $i$ but that solution does not offer more insight. In the second case we can express $i$ and $r$ as a function of the angle $\alpha$ that characterizes the reflection in the mirror's rest frame via the aberration of light formulas[1]

$$\cos i = \frac{\cos\alpha + \frac{V}{c}}{1 + \frac{V}{c}\cos\alpha} \tag{10}$$

and

$$\cos r = \frac{\cos\alpha - \frac{V}{c}}{1 - \frac{V}{c}\cos\alpha} \tag{11}$$

with which (6), (8) and (9) become

$$D_i = \frac{dt_1}{dt_0'} = \frac{\sqrt{1 - \frac{V^2}{c^2}}}{1 + \frac{V}{c}\cos\alpha} \tag{12}$$

$$D_r = \frac{dt_2}{dt_0'} = \frac{\sqrt{1 - \frac{V^2}{c^2}}}{1 - \frac{V}{c}\cos\alpha} \tag{13}$$

$$D = \frac{dt_2}{dt_1} = \frac{1 + \frac{V}{c}\cos\alpha}{1 - \frac{V}{c}\cos\alpha} \tag{14}$$

defining three Doppler factors $D_i$, $D_r$ and $D$.

Figure 4 presents a plot of $D_i$ and $D_r$ as a function of $\alpha$ for different values of $\beta = \frac{V}{c}$. Figure 5 presents a plot of $D$ as a function of $\alpha$ for different values of $\beta = \frac{V}{c}$. In the case of a normal incidence ($\alpha = 0$) we recover Grieser's[4] result

$$D_{\alpha=0} = \frac{1 + \frac{V}{c}}{1 - \frac{V}{c}} \tag{15}$$

a largely used result when deriving relativistic formulas and known as radar echo or police radar. Our approach is a variant of the approach proposed by Gjurchinovski.[5]

Because all the formulas we have used can be derived without using the Lorentz transformations we can consider that our derivation is free of them, stressing the part played by clocks. Involving only changes in the readings of the clocks of the same reference frame, the experiment that leads to (14) does not require clock synchronization at all.

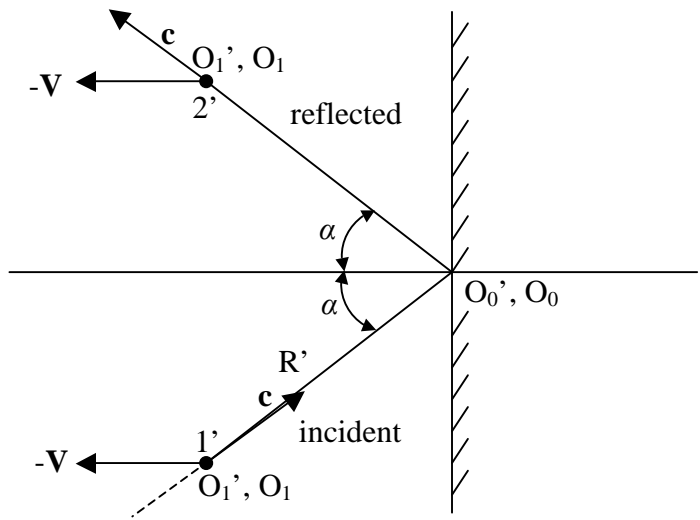

Figure 1. Reflection of light on the stationary mirror.

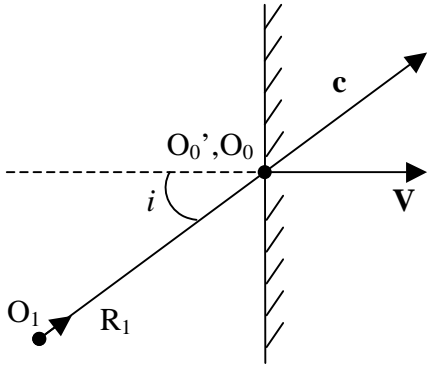

Figure 2. Incidence of light as detected from the reference frame relative to which the mirror moves.

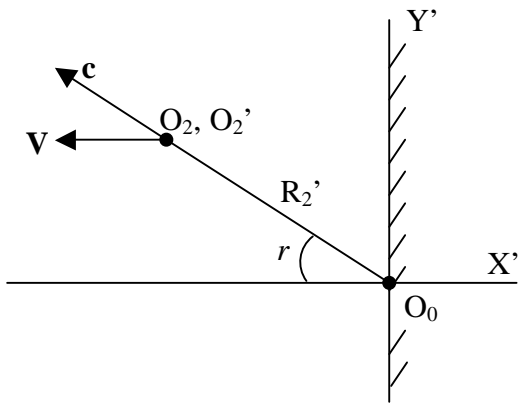

Figure 3. ection of light as detected from the reference frame relative to which the mirror moves.

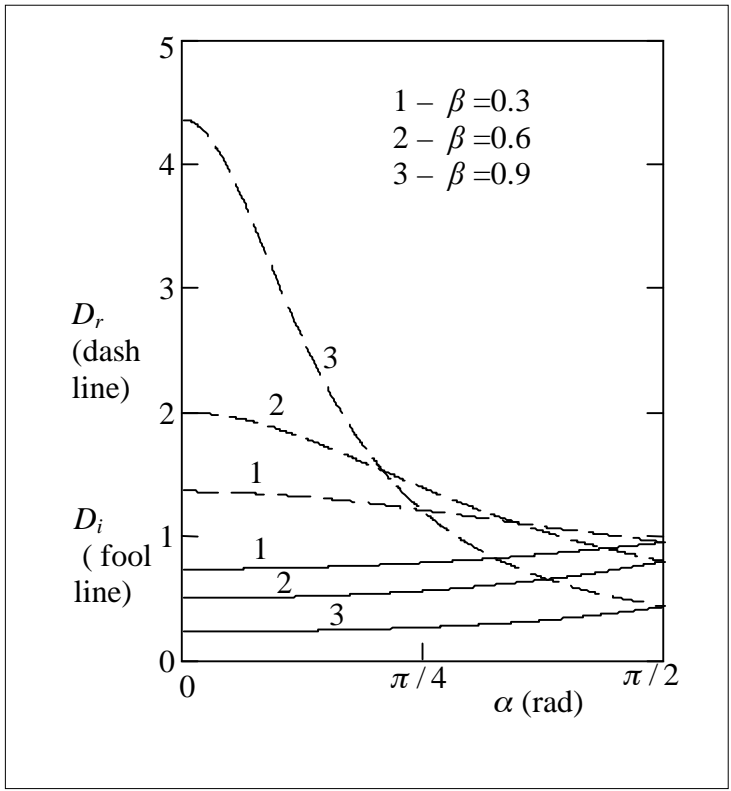

Figure 4. A plot of the Doppler factors $D_i$ and $D_r$ as a function of the angle $\alpha$ that defines the reflection on the mirror in its rest frame for different values of the relative velocity $\beta = V/c$

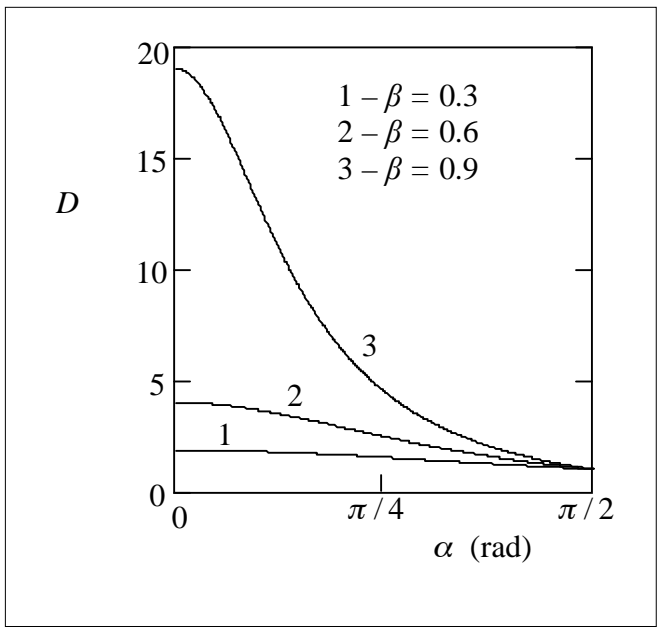

Figure 5. The Doppler factor D that defines the radar echo as a function of the angle $\alpha$ that defines the reflection in the mirrors rest frame for different values of the relative velocity $\beta = V/c$